# Post Mortem System – Playback of the RHIC Collider[*]

J.S. Laster, T. Clifford, T. D'Ottavio, A. Marusic, J.F. Skelly,
Brookhaven National Laboratory, Upton, NY 11973, USA

## Abstract

A Post Mortem System was developed for the Relativistic Heavy Ion Collider at Brookhaven National Laboratory (BNL) to provide a playback of the collider state at the time of a beam abort, quench, or other failure event. Post Mortem data is used to provide diagnostics about the failure and to improve future stores. This data is read from hardware buffers and is written directly to the main file system by Accelerator Device Objects in the front-end computers. The Post Mortem System has facilitated analysis of loss monitor and power supply data, such as beam loss during magnet quenches, dump kicker misfires and power supply malfunctions. System details and recent operating experience will be discussed.

## 1 INTRODUCTION

The post mortem system at RHIC (Relativistic Heavy Ion Collider) was designed to aid physicists and engineers in determining the cause for beam loss or other mechanisms that cause the accelerator to abort the beam in the RHIC machine. One of the major reasons for the necessity of this system is the use of superconducting magnets in the RHIC accelerator. Each magnet has a tolerance level of beam loss that it can accept. Any additional beam loss above the threshold would likely cause a magnet to quench. Not only is this disruptive to the physics program, but damage to the magnet could occur.

There are many systems that are responsible for maintaining beam operations. These include 'beam-control' systems, such as power supply systems, rf systems, vacuum systems, as well as others. Other systems in place perform measurements and diagnostics, such as beam position monitor and wall current monitor systems for beam tuning. The beam loss monitoring system provides beam tuning diagnostics, and hardware and environmental protection.

Any system involved in maintaining beam in the accelerator and diagnosing why beam has been lost from the accelerator may find its place in the post mortem system.

## 2 SYSTEM COMPONENTS

### 2.1 Data Acquisition

The design of the data acquisition system (Fig. 1) and its capability to buffer data lies at the heart of this system. The ability to continuously collect and buffer data, then freeze the buffers due to an external event is what makes the playback of the accelerator state possible. Response times to events and the ability to time-correlate the data to an event are all critical in making this system work. The hardware used to accomplish this task are specially designed MADCs[1] (multiplexed analog to digital converter system) and WFGs (waveform generators).

The MADCs and WFGs can be configured to take some action upon receiving an external event or trigger. For post mortem systems, the action is to immediately stop collecting data or to collect an additional second or more of data prior to freezing the buffer.

FECs (front-end computers) contain software objects called ADOs[2] (accelerator device objects) that abstract the concept of control devices. In the case of post mortem, after a trigger is received, the FEC delivers to the MADC ADO a signal indicating it is time to read the data from the hardware buffers and write the data to disk. A task in the FEC is also signalled to write the WFG ADO data.

### 2.2 Data Storage

The ADOs write data to the file system in the form of SDDS[3] files which are placed in a generic directory. These are immediately post-processed by a console-level application called the post MortemServer to make them ready for final use and to free the generic directory for subsequent reuse. Performing these functions in a server provides flexibility since it is much less disruptive to modify console-level software than front-end software. The functions performed by the postMortemServer include:

- Entry of a database header to indicate that the event was recognized, and that processing of its data is underway.
- Similar data (by system, by timestamp, etc.) are stored together.

---

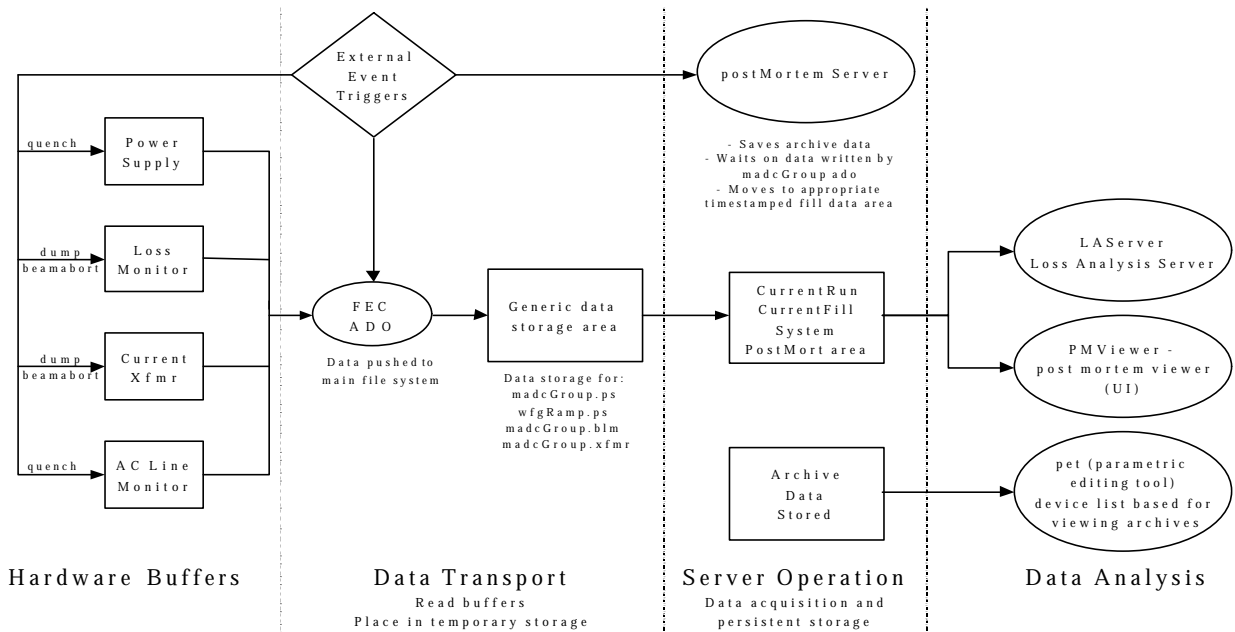

Figure 1. Layout of the Post Mortem System

- Management of dataflow, and coordination of notification to other dependent processes.
- The SDDS files are converted from ascii format to binary to save disk space, since FEC software does not support binary format.
- Spawning processes to save additional device data into archive files, beyond the buffered array data written by ADOs.
- Final revision of the database header description to indicate the current and final status of event preparation, and to indicate that the archives are available for inspection.

### 2.3 User Interface

A user interface was designed to view post mortem data. The viewer allows a user to select a system, find the event of interest, and then display the data about the system. Event of interest selection is assisted by filters, such as run name selection (RHIC data is stored within runs – e.g. rhic_au_fy01) and time limiting options (e.g. day, week, month, entire run). The process reads from the stored data files, places the data into generic post mortem data objects, and then displays the data in graphic form to the user. Basic systems can then have their data displayed with little additional programming.

Reading of data is handled generically for most systems. For systems that have single data sets for objects no additional work is required (e.g. the loss monitor system has one data set per monitor). More complex systems that have natural abstractions, such as the power supply system, require additional effort. A power supply will have many data sets that represent its various properties (current, reference current, voltage, etc.). This abstraction requires an object that comprises many data sets. Utilizing this technique has proved to be very valuable in maintaining core code that allows for system expansion with relative ease.

The user interface has provisions for systems to provide system-specific operations on, and specialized displays of, the data. For example, loss monitor data sets can be searched for areas of high loss to be presented to the user, rather than displaying all of the post mortem data and requiring visual inspection by the user. The loss monitor data also makes use of a 3-D display that provides a user with an overall view of beam loss just prior to and after an event occurs.

### 2.4 Post-processing

One system was developed that makes use of the data automatically, the Loss Analysis Server (LAServer). Each time beam is removed from the collider, an investigation is conducted into the locations in which beam may have been lost. This is accomplished by saving the loss monitor and current transformer post mortem data after an event. The LAServer makes use of the generic post mortem data objects to read the data. It then performs an analysis of the data to reach conclusions about where the beam has been removed from the machine (a beam dump is the desired location for beam to be placed when it is removed from the machine).

## 3 POST MORTEM IN ACTION

### 3.1 Performance

The post mortem system has had its share of growing pains. In its infancy, some data were stale or missing, timestamps were incorrect, files were lost or stranded (not moved to long-term data storage areas), and servers were unable to perform their tasks. The present system has become quite reliable. There are still occasions of improper timestamps resulting in stranded files, in less than 0.5% of the cases.

The data storage requirements differ from system to system. Buffer length, (e.g. 12 seconds of loss monitor data are stored, 4 seconds of power supply data are stored), frequency (e.g. 720Hz for loss monitor and power supply data), and the accuracy of the data all play a role in determining this requirement.

The data saved in a post mortem of power supply data uses approximately 145MB per event (130MB of WFG data, 15MB of MADC data). The loss monitor post mortem data uses an average of 44MB per event. Post mortem data for a typical month requires approximately 120GB of disk space.

Other systems may be affected by the post mortem system. The effect on other data collection systems includes loss of data due to frozen hardware buffers, and the effect of writing a large amount of data to disk, thereby slowing other processes and/or making their data incorrect or obsolete. The current philosophy is that the post mortem data is the most critical.

Data from the beam loss monitor, current transformer, dxheater, power supply, line ac monitoring, and real-time data link monitoring systems have been incorporated into the post mortem system. New systems are continually brought on line as the need arises – and as resources are available.

### 3.2 Examples

Power supply data is analyzed after quench events. The characteristics of the power supplies in the quench location are then studied to determine why the supply failed, if the failure was radiation induced, or if some other type of failure occurred. If the failure is believed to be radiation induced, further analysis of the loss monitor data is conducted. The focus is to determine where and why the beam was lost in order prevent a recurring problem. The amount of beam loss can be determined by studying the entire ring's losses along with the beam current transformer post mortem data.

Some quotes from Operational logs:

Sunday September 30, 2001, 1424 – 'The post mortem plots show voltage ripples for many of the supplies, throughout the ring. Apparently a power glitch had occurred.'

Friday October 6, 2001, 0234 – 'Blue quench link interlock. The post mortem plots showed that the voltage was railed for the bi5-tq6 supply before the link was pulled. … was contacted and reported that the supply needs to be replaced.'

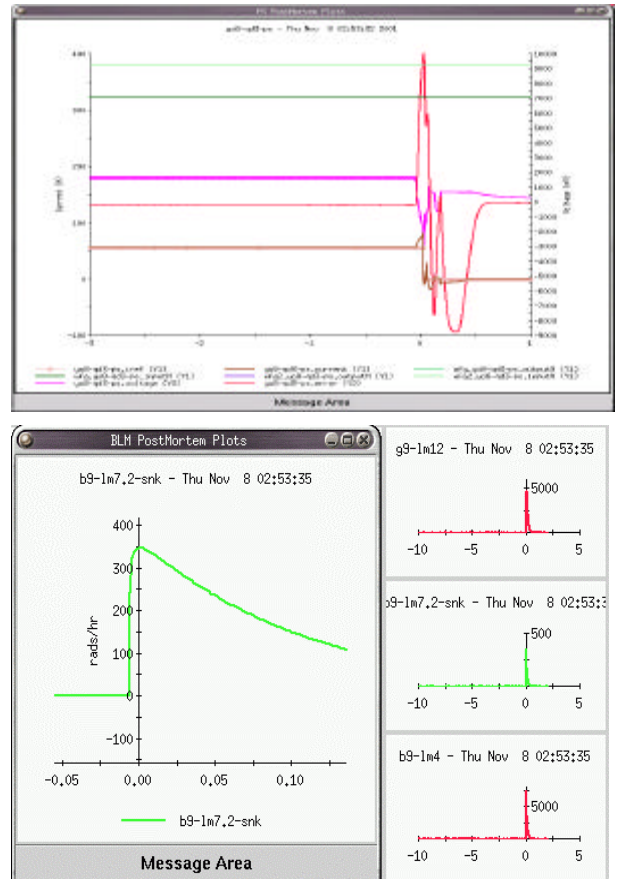

Figure 2. Power Supply and Loss Monitor Data

As shown in Figure 2, a power supply (top) – out of regulation before quench event (time 0 is time of event). Loss monitors (bottom) show beam loss prior to time of event – due to power supply problem – enlarged picture shows loss prior to time of event.

## 4 THE FUTURE

New systems, as appropriate, will be added to the post mortem system. Items to be considered include filtering of event selections at the console level without repeated calls to the database. This will result in improved UI interaction for users of the viewer.